\documentclass[12pt]{article}
\usepackage{latexsym}
\usepackage{amsmath}
\usepackage{amsfonts}
\usepackage{amssymb}
\usepackage{graphicx}

\usepackage{color}
\usepackage{tikz}
\usepgflibrary{arrows}

\newcommand{\mS}{\sum\limits_\lambda\int\frac{d^3\vec{k}}{(2\pi)^32k^0}}
\newcommand{\mSE}{\sum\limits_{\lambda=\pm1}\int\frac{d^3\vec{k}}{(2\pi)^32k^0}}
\newcommand{\mc}{c(\vec{k},\lambda)}
\newcommand{\md}{c^+(\vec{k},\lambda)}
\title{Lorentz transformations, sideways shift and massless spinning particles }
\author{K. Bolonek-\L{}aso\'n$^1$, P. Kosi\'nski$^2$, P. Ma\'slanka$^2$\thanks{e-mail: pmaslan@uni.lodz.pl}\\
$^1$\small Department of Statistical Methods
 Faculty of Economics and Sociology\\
$^2$\small Department of  Computer Science, 
 Faculty of Physics and Applied Informatics\\
\small University of \L\'od\'z,\\
\small Pomorska 149/153, 90-236 {\L}\'od\'z, Poland}

\date{}
\begin{document}
\maketitle
\begin{abstract}
Recently (Phys. Rev. Lett. \textbf{114} (2015), 210402) the influence of the so called "Wigner translations" (more generally-Lorentz transformations) on circularly polarized Gaussian packets ( providing the solution to Maxwell equations in paraxial approximation) has been studied. 
 It appears that, within this approximation, the Wigner translations have an effect of shifting the wave packet trajectory parallel to itself by an amount proportional to the photon helicity. It has been suggested that this shift may result from specific properties of the algebra of Poincare generators for massless particles. 
 In the present letter we describe the general relation between transformation properties of electromagnetic field on quantum and classical levels. It allows for a straightforward derivation of the helicity-dependent transformation rules. We present also an elementary derivation of the formula for sideways shift based on classical Maxwell theory. Some comments are made concerning the generalization to higher helicities and the relation to the coordinate operator defined long time ago by Pryce.
 
\end{abstract}

\newpage
\section{ Introduction}
The issue of Poincare covariance has been a subject of intensive study for many years, starting from the seminal paper of Wigner \cite{b1}. Since then there have appeared numerous papers devoted to the various aspects 
of this problem.
\par
Recently, a renewed interest in this topic has been observed which is related to the problem of Lorentz covariance \cite{b2}, \cite{b3}, \cite{b4} of chiral kinetic theory with anomalous  conservations laws \cite{b5}, \cite{b6}, \cite{b7}, localization of massless particles \cite{b8} and the Hall effect of light \cite{b9}, \cite{b10}, \cite{b11}. These and  related topics where further studied in Refs. \cite{b12}$\div$\cite{b15}.

\par
 In the recent interesting paper  Stone et al. \cite{b16} analyzed the role of Wigner translations in transformation properties of finite-size wave packets of non-zero helicity ( circularly polarized ). It appeared that Wigner translations result in sideways shift of the wave packet trajectory. 
More specifically, the authors of Ref. \cite{b16} considered an explicit example of circularly polarized Gaussian beam in the paraxial approximation to Maxwell equations. They computed the sideways shift of energy density and energy flux under Lorentz transformations.The actual calculations appear to be rather complicated but the final result is quite simple and transparent. It has been already argued in \cite{b2} that a similar shift occurs in the case of Lorentz transformations applied to massless particles of non-zero helicity. Stone et. al. suggested that the latter can be explained by a simple algebraic argument involving the algebra of Poincare generators and, moreover, both phenomena are related.

\par
In the present paper we analyse the problem from more general point of view. We show that the sideways shift resulting from Lorentz transformations of massless particles carrying non-zero helicity is closely related to the one computed by Stone et al.  and the whole effect is a direct consequence of standard properties of unitary representations of Poincare group. 
\par
Our starting point is the description of such representations for massless particles of arbitrary helicity. We remind the explicit form of Poincare generators in the single particle theory. Due to the irreducibility of the representations under consideration any observable can be, at least in principle, constructed in terms of Poincare generators . In particular, one can define the coordinate operators which allow to rewrite the Poincare generators in a simple and transparent way ( cf. eqs. (\ref{e6}) below). The commutation relations between Poincare generators and coordinate operator determine the transformation properties of the latter. In the particular case of electromagnetic field ( helicity one) it is not difficult to relate the expectation values of coordinate operator to certain classical quantities. To this end we consider a coherent state of electromagnetic field corresponding to the momentum profile strongly peaked at some wave vector. The expectation value of coordinate operator in such a state equals (up to a scalar factor) the energy density centroid computed from corresponding classical field configuration (see eq. (\ref{e20}) below). The sideways shift of the latter resulting from Lorentz transformations can be, in turn, computed in an elementary way using classical Maxwell equations. It coincides with the expression obtained from transformation properties of helicity one massless particles.
\par 
We conclude the paper with some remarks concerning the generalization of the above results to higher helicities and the relationship with coordinate operator introduced long time ago by Pryce \cite{b17}.

\section{Massless particles with arbitrary helicities}

 As it has been explained by numerous authors (starting from Ref. \cite{b1}) massless particle carrying helicity $\lambda$ is described by an unitary representation of Poincare group induced from the homomorphic representation of stability subgroup of the standard fourvector (say) $\underline{k}^\mu=(\underline{k},0,0,\underline{k})$. The stability subgroup is isomorphic to the group $E(2)$\ of rigid motions of Euclidean plane and the kernel of the representation of the latter used for description of massless particles consists of two translations in the plane ("Wigner transformations"). The resulting induced representation of Poincare group is characterized by a single (half)integer quantum number $\lambda$ called helicity. The Poincare  generators $P_\mu$ (translations) and $M_{\mu\nu}$\ (Lorentz transformations) read  \cite{b18} 
\begin{equation}
\label{e1}
\begin{split}
&\hat{H}\equiv P^0 = k^0=\mid \vec{k} \mid   \\
&\hat{\vec{P}} = \vec{k}  \\
&\hat{\vec{M}} = \vec{k} \times (-i\vec{\nabla}_k)+\lambda\vec{m }(\vec{k}) \;\;\;(M^i\equiv \frac{1}{2}\varepsilon_{ijk}  M^{jk})  \\
&\hat{\vec{N}} = ik^0\vec{ \nabla}_k+\lambda\vec{n}(\vec{k}) \;\;\;\;\;(N^i\equiv M^{i0})
\end{split}
\end{equation}
where $\vec{m}(\vec{k})$\ and $\vec{n}(\vec{k})$\ are given by
\begin{equation}
\label{e2}
\begin{split}
&\vec{m}(\vec{k}) = \frac{\vert\vec{k}\vert}{(k^1)^2 + (k^2)^2} (k^1,k^2,0) \\
&\vec{n}(\vec{k}) = \frac{k^3}{(k^1)^2 + (k^2)^2} (-k^2,k^1,0)  \\
\end{split}
\end{equation}
  The generators act in the Hilbert space of functions $f(\vec{k},\lambda)$ ($\lambda$-fixed) equipped with the scalar product 
\begin{equation}
\label{e3}
(f,g)  \equiv \int\frac{d^3\vec{k}}{(2\pi)^32k^0}\overline{f(\vec{k},\lambda)}g(\vec{k},\lambda)
\end{equation}
Due to the irreducibility of the representation under consideration, any operator acting in our Hilbert space can be, in principle, constructed from Poincare generators (modulo domain problems). In particular, one can construct coordinate operator $\hat{\vec{x}}=(\hat{x}^1,\hat{x}^2,\hat{x}^3)$ as follows
\begin{equation}
\label{e4}
\hat{\vec{x}} = i\vec{\nabla}_k - \frac{i}{2}\frac{\vec{k}}{(k^0)^2}+\lambda\frac{\vec{n}(\vec{k})}{k^0 }; 
\end{equation}

They obey the following commutation rules
\begin{equation}
\label{e5}
\begin{split}
&\left[ \hat{x}^i,\hat{x}^j\right] = -i\lambda\varepsilon_{ijk} \frac{k^k}{(k^0)^3}  \\
&\left[\hat{x}^i,k^j\right] = i\delta_{ij}
\end{split}
\end{equation}
Moreover, the generators of Lorentz group can be expressed in terms of $\hat{\vec{x}}$\ and $\vec{k}$ as follows

\begin{equation}
\label{e6}
\begin{split}
&\hat{M}^i = \varepsilon_{ijl} \hat{x}^jk^l + \lambda \frac{k^i}{k^0}  \\
&\hat{N}^i = \frac{1}{2} \left( \hat{x}^ik^0 + k^0\hat{x}^i\right) 
\end{split}
\end{equation}

This form of Poincare generators for massless particles with helicity $\lambda$\ has been proposed long time ago by Atre et al. \cite{b19} and Skagerstam \cite{b20}. Note that it can be also obtained by a straightforward quantization of classical Hamiltonian system defined on appropriate coadjoint orbit of Poincare group \cite{b21}$\div$\cite{b27}, \cite{b4}.
The first, non-standard, commutation rule (\ref{e5}) results from the fact that the original Darboux coordinates defined in the framework of the orbit method are not explicitly $SO(3)$ covariant. Due to the fact that the choice of standard fourvector $\underline{k}^\mu$\  breaks rotational invariance they transform nonlinearly under rotations \cite{b4}. Linearization of $SO(3)$ action yields new coordinates $\hat{\vec{x}}$\  with non-standard Poisson brackets/commutations rules.\\
Passing to the many particle theory yields the following structure. 
 First, we have the creation/annihilation operators obeying the commutation rules

\begin{equation}
\label{e7}
\left[ c(\vec{k},\lambda),c^+(\vec{k'},\lambda')\right] = (2\pi)^32k^0\delta^3(\vec{k}-\vec{k'})\delta_{\lambda\lambda'}
\end{equation}
 where we have admitted both helicities .\\
The momentum and boost generators  read
\begin{equation}
\label{e8}
\begin{split}
&\hat{\vec{M}} = \mS \md \left( \vec{k}\times(-i\vec{\nabla}_k) + \lambda\vec{m}(\vec{k})\right) \mc \\
&\hat{\vec{N}} =\mS c^+(\vec{k},\lambda,t) \left(ik^0\vec{\nabla}_k - t\vec{k} + \lambda\vec{n}(\vec{k})\right) c(\vec{k},\lambda,t)
\end{split}
\end{equation}
where $c(\vec{k},\lambda,t) = exp(-ik^0t)\mc$\ and $c^+(\vec{k},\lambda,t) = exp(ik^0t)\md$\ are the annihilation and creation operators in the Heisenberg picture. The explicit forms  of the remaining generators are not needed here.

\section{The electromagnetic field}

Consider now the case of electromagnetic field, $\mid \lambda\mid = 1$. The relevant field operator reads  \cite{b18} :

\begin{equation}
\label{e9}
\begin{split}
&\hat{F}_{\mu\nu}(x) = 
 \mSE \left( e_{\mu\nu}(\vec{k},\lambda)e^{i\vec{k}\vec{x}} c(\vec{k},\lambda,t) + \overline{e_{\mu\nu}(\vec{k},\lambda)}e^{-i\vec{k}\vec{x}} c^+(\vec{k},\lambda,t) \right)
\end{split}  
\end{equation}

where $e_{\mu\nu}(\vec{k},\lambda) = \overline{e_{\mu\nu}(\vec{k},-\lambda)}$\ are the appropriate polarization tensors and $x$\ denotes c-number space-time coordinates. 

The generators (\ref{e8}) can be expressed in terms of energy-momentum tensor  $\hat{T}^{\mu\nu} = :\hat{F}^\mu_{\;\;\;\alpha}\hat{F}^{\nu\alpha}: - \frac{1}{4}g^{\mu\nu}:\hat{F}_{\alpha\beta}\hat{F}^{\alpha\beta}:$

\begin{equation}
\label{e10}
\begin{split}
&\hat{P}^k = \int d^3  \vec{x }\;\;\hat{T}^{k0}(x) \\
&\hat{N}^k = \int d^3  \vec{x }\;\;\left( x^k\hat{T}^{00}(x) - t \hat{T}^{k0}(x) \right)
\end{split}   
\end{equation}
leading to 

\begin{equation}
\label{e11}
 \int d^3  \vec{x }\;\; x^k \hat{T}^{00}(x)    = \hat{N}^k + t\hat{P}^k
\end{equation}

Using eqs. (\ref{e6}), (\ref{e8}) and (\ref{e11}) one finds  

\begin{equation}
\label{e12}
 \int d^3  \vec{x }\;\; x^i\hat{T}^{00}(x)    = \mSE c^+(\vec{k},\lambda,t) \frac{1}{2}\left( \hat{x}^ik^0 + k^0\hat{x}^i\right) c(\vec{k},\lambda,t)
\end{equation}
Eq.(\ref{e12}) is an identity which allows us to relate the expectation value of the coordinate operator to energy-density centroid of classical electromagnetic field. To this end we define the second-quantized version of coordinate operator as

\begin{equation}
\label{e13}
 \hat{\vec{X}}(t)   = \mSE c^+(\vec{k},\lambda,t)\; \hat{\vec{x}}(i\vec{\nabla}_k,\vec{k},\lambda)\;c(\vec{k},\lambda,t)
\end{equation}

Consider now a coherent states describing field configuration of definite helicity and profile $f(\vec{k})$ : 

\begin{equation}
\label{e14}
\begin{split}
&|f \rangle \equiv exp\left( \int\frac{d^3\vec{k}}{(2\pi)^32k^0} \left( f(\vec{k},\lambda) \md-\overline{f(\vec{k},\lambda)} \mc \right) \right) |0\rangle = \\
&=exp\left( -\frac{1}{2}\int\frac{d^3\vec{k}}{(2\pi)^32k^0} |f(\vec{k},\lambda)|^2\right) exp\left( \int\frac{d^3\vec{k}}{(2\pi)^32k^0} f(\vec{k},\lambda) \md\right)  |0 \rangle
\end{split}
\end{equation}
Note that

\begin{equation}
\label{e15}
 c(\vec{k},\lambda')|f \rangle = \delta_{\lambda\lambda'}f(\vec{k},\lambda) |f \rangle
\end{equation}

The classical field corresponding to the above coherent state reads

\begin{equation}
\label{e16}
\begin{split}
&F_{\mu\nu}^{clas}(x) \equiv \left\langle f \mid \hat{F}_{\mu\nu}(x) \mid f\right\rangle =  \\
&\;\;\;\;\;\;\;\; =\int\frac{d^3\vec{k}}{(2\pi)^32k^0} \left( e_{\mu\nu}(\vec{k},\lambda)e^{-ikx} f(\vec{k}) + \overline{e_{\mu\nu}(\vec{k},\lambda)}e^{ikx} \overline{f(\vec{k})} \right) 
\end{split}
\end{equation}
Now , eqs. (\ref{e8}), (\ref{e11}), (\ref{e15}) imply
\begin{equation}
\label{e17}
\int d^3\vec{x} x^i\left\langle f \mid \hat{T}^{00}(x)\mid f \right\rangle = \int d^3\vec{x} \;x^i \;T^{00}(F^{clas}(x))
\end{equation}
In fact, using eq.~(\ref{e15}) we find that taking the 
expectation value of any normally ordered bilinear form in 
creation and annihilation operators is equivalent to replacing 
the relevant operators by the profile $f(\vec{k})$ and its 
complex conjugate. On the other hand, the same expression is 
obtained by computing its classical counterpart for the field 
configuration (\ref{e16}).
By virtue of eqs. (\ref{e12}), (\ref{e15}) and (\ref{e17}) one can write
\begin{equation}
\label{e18}
\begin{split}
&\int d^3\vec{x} \;x^i \;T^{00}(F^{clas}(x))=\\
 &\int\frac{d^3\vec{k}}{(2\pi)^32k^0}\overline{f(\vec{k})}e^{ik^0t}\frac{1}{2}(\hat{x}^ik^0+k^0\hat{x}^i)e^{-ik^0t}f(\vec{k})
\end{split}
\end{equation}

For the profiles strongly peaked at some wave vector the right hand side approximately factorizes into

\begin{equation}
\label{e19}
\begin{split}
&\frac{1}{\|f\|^2}\int\frac{d^3\vec{k}}{(2\pi)^32k^0}k^0 \mid f(\vec{k})\mid^2 \cdot \int\frac{d^3\vec{k}}{(2\pi)^32k^0} 
\overline{f(\vec{k})}e^{ik^0t}\hat{x^i}e^{-ik^0t}f(\vec{k})= \\
&\frac{1}{\|f\|^2}\left\langle f \mid \hat{H} \mid f\right\rangle \left\langle f \mid \hat{X}^i(t) \mid f\right\rangle =
\frac{1}{\|f\|^2}E(F^{clas})\left\langle f \mid \hat{X}^i(t) \mid f\right\rangle
\end{split}
\end{equation}
where $\|f\|^2=\int\frac{d^3\vec{k}}{(2\pi)^32k^0} \mid f(\vec{k})\mid^2 $. Eqs.(\ref{e18}) and (\ref{e19}) imply then

\begin{equation}
\label{e20}
\frac{\left\langle f \mid \hat{X}^i(t) \mid f\right\rangle}{\|f\|^2} {\sim} \frac{\int d^3\vec{x} \;x^i \;T^{00}(F^{clas}(x))}{\int d^3\vec{x}  \;T^{00}(F^{clas}(x))}
\end{equation}

Eq.(\ref{e20}) relates the expectation value of second quantized coordinate operator in coherent state of electromagnetic field to energy density centroid of the corresponding classical field configuration. Since ${\|f\|^2}$\ is a scalar factor the transformation rules of both quantities should coincide. The transformation rule of the left hand side of eq.(\ref{e20}) follows from the properties of the relevant unitary representation of Poincare group. On the other hand, the transformation properties of the right hand side can be easily derived using classical Maxwell theory. Some details will be given in the next section.

\section{Lorentz transformation }

Let us start with the transformation properties of energy density centroid. For a given configuration of electromagnetic field vanishing sufficiently fast at spatial infinity to justify all integrations by parts necessary to derive the formulae given below we define

\begin{equation}
\label{e21}
\begin{split}
&E \equiv  P^0\equiv\int d^3\vec{x}\;T^{00}(x) \\
&P^k \equiv \int d^3\vec{x}\;T^{k0}(x) \\
&M^{ik} \equiv \int d^3\vec{x}\;\left(x^iT^{k0}(x)-x^kT^{i0}(x)\right) \\
&\mathcal{X}^{k} \equiv \frac{1}{E} \int d^3\vec{x}\;x^kT^{00}(x)
\end{split}
\end{equation}
Then $E=const$, $\vec{P}=const$\ and

\begin{equation}
\label{e22}
\mathcal{\dot X}^k=\frac{P^k}{E}
\end{equation}
Consider now an infinitesimal Lorentz transformation, $\Lambda^\mu_{\;\;\nu}=\delta^\mu_{\;\;\nu}+\omega^\mu_{\;\;\nu},\;\; \omega^{\mu\nu}=-\omega^{\nu\mu}$. Using $ x'^\mu=x^\mu+\omega^\mu_{\;\;\nu}x^\nu,\;\;T'^{\mu\nu}(x')=T^{\mu\nu}(x)+\omega^\mu_{\;\;\alpha}T^{\alpha\nu}(x)+\omega^\nu_{\;\;\alpha}T^{\mu\alpha}(x)$, expanding everything to first order in $\omega^\mu_{\;\;\nu}$, using the continuity equation for $T^{\mu\nu}$\ and integrating by parts we easily find from eq.(\ref{e20})
\begin{equation}
\label{e23}
\mathcal{X'}^i=\mathcal{X}^i+\omega^i_{\;\;k}\mathcal{X}^k+
\omega^i_{\;\;0}x^0-\omega^0_{\;\;k}\frac{\mathcal{X}^iP^k}{E}
+\omega^0_{\;\;k}\frac{M^{ik}}{E}
\end{equation};

here $x^0=t$ is the time coordinate. Let us compare eq.(\ref{e23}) with the transformation rule for a trajectory of free massless point particle, $y^\mu=y^\mu(y^0)$. Performing an infinitesimal Lorentz transformation and taking into account the correction due to the change of time variable we find: $y'^i=y^i+\omega^i_{\;\;k}y^k+\omega^i_{\;\;0}y^0-\omega^0_{\;\;k}y^i\frac{\pi^k}{\varepsilon}$, where $\pi^i$\ and $\varepsilon$\ are the momentum and energy, respectively. Therefore, one can rewrite eq. (\ref{e23}) in the form

\begin{equation}
\label{e24}
\mathcal{X}'^i=\mathcal{X}^i+\omega^i_{\;\;k}\mathcal{X}^k+\omega^i_{\;\;0}\mathcal{X}^0-\omega^0_{\;\;k}\mathcal{X}^k\frac{P^i}{E}+\omega^0_{\;\;k}\frac{M^{ik}-(\mathcal{X}^iP^k-\mathcal{X}^kP^i)}{E}
\end{equation}
and identify the last term on the right hand side as the contribution from spin-dependent sideways shift. For a definite helicity and profile $f(\vec{k})$\ strongly peaked at some wave vector we find the following expression for sideways shift 
 
\begin{equation}
\label{e25}
\delta \mathcal{X}^i\sim\lambda\;\varepsilon_{ikl}\;\omega^0_{\;\;k}\frac{P^l}{E^2}
\end{equation}

The transformation rules in classical theory are compatible with those obtained on quantum level. Indeed, on the one particle level one finds from eqs. (\ref{e5}) and (\ref{e6})

\begin{equation}
\label{e26}
\begin{split}
&\left[ \hat{N}^i,\hat{x}^l\right]=\left[ \frac{1}{2}\left( \hat{x}^ik^0+k^0\hat{x}^i\right)-tk^i,\hat{x}^l\right] = \\
&-i\left( \hat{M^{il}}-(\hat{x}^i(t)k^l-\hat{x}^l(t)k^i)\right) - \frac{i}{2}\left(\hat{x}^i\frac{k^l}{k^0} +\frac{k^l}{k^0}\hat{x}^i\right) +it\delta_{il} 
\end{split}
\end{equation}

and it is not difficult to verify  that the  transformation properties of the left hand side of eq. (\ref{e20}) agree with  those of the right hand side.

\section{Conclusions }

For any classical electromagnetic field configuration such that the relevant integrals are convergent one can define the energy density centroid $\mathcal{\vec{X}}$.  The Lorentz transformation rule of $\mathcal{\vec{X}}$\ follows easily from continuity equation for energy-momentum tensor. It takes particularly simple form for field configuration corresponding to momentum profiles strongly peaked at some wave vector. The centroid of the circularly polarized wave transforms as a coordinate of free massless point particle plus an additional term (sideways shift) equipped with the sign depending on the direction of polarization (eq.(\ref{e25}).
\par
On the quantum level we are dealing with massless particles carrying non-zero helicity (helicity one in the case of electromagnetic field). They are described by the irreducible representations of Poincare group induced from homomorphic representations of $E(2)$\ subgroup. Within  such a representation one can construct the coordinate operator. Poincare generators, when expressed in terms of the latter, take a particularly simple form. On the other hand,the classical field configurations are described by coherent states. Therefore, to make contact with classical description one has pass to many particle description.This is done in a standard way using the formalism of second quantization. The coordinate operator is now described by eq.(\ref{e13}) and one can compute its expectation value in coherent state of definite helicity. It appears that for the profile $f(\vec{k})$\ strongly peaked at same $\vec{k}$ there exists a simple relation between the expectation value of coordinate operator and the classical
energy centroid (eq. (\ref{e20})). It suggests that the transformation properties of the classical centroid can be also derived from the algebra of generators of the relevant unitary representation. The normalization factor on the left hand side of eq. (\ref{e20}) calls for some comment. Its origin has a simple explanation. The many particle coordinate operator is not canonically conjugated to the total momentum $\hat{P}$\ defined by eq. (\ref{e8}).
In fact their commutator reads  

\begin{equation}
\label{e27}
\begin{split}
&\left[ \hat{X}^i(t),\hat{P}^k\right]= i \delta_{ik}\sum_\lambda \int\frac{d^3\vec{k}}{(2\pi)^32k^0}c^+(\vec{k},\lambda)c(\vec{k},\lambda)=i\delta_{ik}\mathcal{\hat{N}}
\end{split}
\end{equation}
where $\mathcal{\hat{N}}$\ is the photon number operator.\\
The normalization factor $\|f\|^2= \langle f |\mathcal{\hat{N}}|f\rangle$\ is the expectation value of the number of photons. Roughly speaking, the expectation value of $\hat{X}^i(t)$\ is (averaged) sum of coordinates of all photons. The "center of mass " coordinate should be obtained by dividing the sum of coordinates by the number of photons. However, this cannot be done on the operator level because $\mathcal{\hat{N}}$\ is not invertible; were it not so it would be possible to define center of mass coordinate canonically conjugated to total momentum. On the other hand, eq. (\ref{e20}) can be viewed as the equality of energy density centroid and the expectation value of center of mass coordinate for particular class of coherent states.
\par
Let us note that the coordinate operator considered here is related to the one discussed long time ago by Pryce \cite{b17} who analyzed a number of possibilities of defining the relativistic coordinate.

The question arises if the above reasoning can be extended to higher helicities, $|\lambda|>1$. The construction involving Poincare generators is general and the resulting formulae are valid for any $\lambda$. However, the crucial point is the existence of energy-momentum tensor $T^{\mu\nu}$. By Weinberg-Witten theorem \cite{b28} in the case of massless particles such a tensor exists only provided $|\lambda|\leq1$\ and the reasoning presented here cannot be extended to higher helicities

{\bf Acknowledgements}\\
We are very grateful to Cezary Gonera, Joanna Gonera and Krzysztof Andrzejewski for interesting discussion. 
P.K. and P.M. were supported by National Science Center, Poland, grant number DEC-2013/09/B/ST2/02205

\end{document}